\documentclass[conference]{IEEEtran}
\IEEEoverridecommandlockouts
\usepackage{cite}
\usepackage{amsmath,amssymb,amsfonts}
\usepackage{algorithmic}
\usepackage{graphicx}
\graphicspath{{figures/}}
\usepackage{textcomp}
\usepackage{xcolor}
\usepackage{hyperref}
\usepackage{longtable}
\def\BibTeX{{\rm B\kern-.05em{\sc i\kern-.025em b}\kern-.08em
    T\kern-.1667em\lower.7ex\hbox{E}\kern-.125emX}}
\begin{document}

\title{Investigating ChatGPT's Potential to Assist in Requirements Elicitation Processes\\
}

\author{\IEEEauthorblockN{Krishna Ronanki\IEEEauthorrefmark{1},
Christian Berger\IEEEauthorrefmark{2} and Jennifer Horkoff\IEEEauthorrefmark{3}}
\IEEEauthorblockA{Dept. of Computer Science and Engineering,
University of Gothenburg\\
Gothenburg, Sweden\\
\IEEEauthorrefmark{1}krishna.ronanki@gu.se,
\IEEEauthorrefmark{2}christian.berger@gu.se,
\IEEEauthorrefmark{3}jennifer.horkoff@gu.se}}

\maketitle

\begin{abstract}
Natural Language Processing (NLP) for Requirements Engineering (RE) (NLP4RE) seeks to apply NLP tools, techniques, and resources to the RE process to increase the quality of the requirements. There is little research involving the utilization of Generative AI-based NLP tools and techniques for requirements elicitation. In recent times, Large Language Models (LLM) like ChatGPT have gained significant recognition due to their notably improved performance in NLP tasks. To explore the potential of ChatGPT to assist in requirements elicitation processes, we formulated six questions to elicit requirements using ChatGPT. Using the same six questions, we conducted interview-based surveys with five RE experts from academia and industry and collected 30 responses containing requirements. The quality of these 36 responses (human-formulated + ChatGPT-generated) was evaluated over seven different requirements quality attributes by another five RE experts through a second round of interview-based surveys. In comparing the quality of requirements generated by ChatGPT with those formulated by human experts, we found that ChatGPT-generated requirements are highly Abstract, Atomic, Consistent, Correct, and Understandable. Based on these results, we present the most pressing issues related to LLMs and what future research should focus on to leverage the emergent behaviour of LLMs more effectively in natural language-based RE activities.
\end{abstract}


\section{Introduction} \label{sec:int}
Due to the increasing access to the huge volumes of data generated all over the world~\cite{iyer2021ai}, there is a growing involvement of AI in our daily lives. Because of this trend, AI systems need to be not only safe and reliable but also trustworthy since they have the potential to directly or indirectly cause harm to users and society~\cite{kaur2022Trustworthy}. Trustworthy AI can be defined as a conceptual framework that ensures that the development and implementation of technically and socially robust AI systems adhere to all the applicable laws and regulations, and conform to general ethical principles~\cite{aihleg}. The importance of Trustworthy AI in the contemporary world cannot be understated, and the forthcoming mandatory compliance with the European Union's AI Act (AIA) guidelines while developing and implementing AI systems underscores its significance.

Requirements Engineering (RE) is considered a critical juncture of the interplay between ethics and technology~\cite{9218169}. The RE process at the beginning of a product development life cycle fosters increased communication and collaboration between various stakeholders, offering opportunities to discuss ethical concerns~\cite{kostova2020interplay}, like aspects associated with the trustworthiness of AI, and incorporate them into the development process in a concrete manner. In the field of RE, the quality of the requirements gathered is a fundamental concern. Several researchers and standards organizations have identified a set of quality attributes that are crucial for RE based on the IEEE standards for requirements specification~\cite{standards2016ieee}.
 
Over the years, empirical evidence has suggested that using natural language is the most prevalent approach for writing requirements in industrial practice. This strong interrelation between natural language and requirements led to the outset of Natural Language Processing (NLP) for RE~\cite{zhao2021natural}. NLP4RE seeks to apply NLP tools, techniques, and resources to the RE processes to support human analysts in carrying out various tasks on textual requirements such as detecting and improving language issues, among other things~\cite{zhao2021natural}, which increase the quality of the requirements. Natural Language Generation (NLG) is a process that we use to generate meaningful phrases, sentences, and paragraphs in natural human language~\cite{khurana2022natural}, and is considered one of the most critical yet complex sub-fields of NLP~\cite{ji2023survey}. But the utilization of Generative AI-based NLP tools and techniques for eliciting requirements in support of RE activities is lacking~\cite{zhao2021natural}, indicating a gap in the current state-of-the-art in NLP4RE. On one hand, if this gap is addressed, it could potentially lead to improvements in the overall quality of artifacts and processes involved in RE.

On the other hand, Large Language Models (LLM) have gained significant recognition due to their notably improved performance in NLP tasks~\cite{haque2022think} in recent times. ChatGPT, based on the GPT-3.5 language model, is optimized for dialogue~\cite{zhang2022would} and is capable of answering questions in a human-like text while keeping track of the entire conversation~\cite{10062688}. Despite being trained on a large general domain data and specifically fine-tuned for conversational tasks~\cite{shen2023chatgpt}, it has been observed to perform surprisingly well on specific technical tasks~\cite{choi2023chatgpt}. The use of AI-based conversational chatbots for critical software development activities is not unprecedented either. Machine Learning (ML) models have been proven to provide multiple advantages while implemented in RE for developing privacy-aware systems~\cite{lee2019confident}. Furthermore, sentiment analysis of Twitter data for early adopters of ChatGPT showed a positive sentiment of 83\% towards enhancing NLP-based tasks~\cite{haque2022think}.

Considering the small number of studies involving requirements elicitation using AI-based models and the emergence of LLMs like ChatGPT that are proficient in interacting using natural language, we seek to investigate the potential of ChatGPT (Feb 9 Version) in the requirements elicitation processes by assessing the quality of requirements obtained using ChatGPT in a controlled context and compare them against requirements formulated by human RE experts. We chose to assess ChatGPT's capacity to generate requirements for a fairly novel and uncharted field, in comparison to human capabilities. The topic of Trustworthy AI has been chosen for investigation as it is an area of significant importance, yet there exists a lack of comprehensive understanding of this subject. To that end, the study seeks to answer the following research questions:

\vspace{0.2cm}
\noindent\fbox{%
\begin{minipage}{.96\columnwidth}
\textbf{RQ1- On which requirements quality attributes did ChatGPT-generated requirements score the highest? }
\end{minipage}
}\\

\vspace{0.2cm}
\noindent\fbox{%
\begin{minipage}{.96\columnwidth}
\textbf{RQ2- What are the identified shortcomings of ChatGPT in generating requirements?}
\end{minipage}
}\\

\vspace{0.2cm}
\noindent\fbox{%
\begin{minipage}{.96\columnwidth}
\textbf{RQ3- How do quality attribute scores of requirements generated by ChatGPT compare with requirements formulated by human RE experts?}
\end{minipage}
}\\

Section \ref{sec:rw} provides relevant background and motivates the research goals. Section \ref{sec:methods} describes the design of our study. Section \ref{sec:results} presents the results of the methods employed and an analysis of how requirements generated by ChatGPT compare to those formulated by human experts, based on the evaluation scores. We discuss the results and present any identified validity threats in Section \ref{sec:discussion} and conclude in Section \ref{sec:conclusion}. 

\section{Related Work} \label{sec:rw}

\subsection{Trustworthy AI qualities}
Kaur et al.~\cite{kaur2021requirements} identified several key AI qualities for Trustworthy AI based on the guidelines proposed by the European Union (EU)~\cite{AIact}, including Accuracy \& Robustness, Safety, Non-discrimination, Transparency \& Explainability, Accountability, Privacy \& Security, Regulations, and Human Agency \& Oversight. The requirements elicitation questionnaire presented in~\ref{list:rtaiqs} was crafted based on the Trustworthy AI qualities presented in this study.

\subsection{Requirements Quality Attributes}
 We selected 7 requirement quality attributes presented by Denger et al.~\cite{denger2005quality} and Genova et al.~\cite{genova2013framework} to evaluate the quality of the requirements gathered through the interview-based surveys and ChatGPT. These attributes include Abstraction, Atomicity, Consistency, Correctness, Unambiguity, Understandability, and Feasibility. 

\subsection{State-of-the-art in LLMs' Application in Software Engineering Activities}
In recent years, transformer-based models such as BERT achieved state-of-the-art results in natural language processing (NLP) tasks. These models are typically pre-trained on large amounts of textual data and then fine-tuned on task-specific data to perform particular NLP tasks. In their study, Mosel et al.~\cite{von2022validity} compare BERT transformer models trained on software engineering (SE) data (context-specific specialized vocabulary) with those trained on general domain data in multiple dimensions: their vocabulary, their ability to understand which words are missing, and their performance in classification tasks. Their results demonstrate that for tasks that require an ``understanding'' of the SE context, pre-training with SE data is valuable yet for general language understanding tasks within the SE domain, models trained on general domain data are sufficient.

Alhoshan et al.~\cite{alhoshan2023zero} report an extensive study using the contextual word embedding-based zero-shot learning approach for requirements classification. The study tested this approach by conducting more than 360 experiments using 4 language models with a total of 1020 requirements and found generic language models trained on general-purpose data perform better than domain-specific language models under the zero-shot learning approach.

Ahmed et al.~\cite{ahmad2023towards} investigate the potential and limitations of using ChatGPT to assist an architect in Architecture-centric Software Engineering (ACSE) using a Human-DevBot collaboration approach. They presented a case study in which a novice software architect collaborates with ChatGPT to analyze, synthesize, and evaluate a services-driven software application. They investigated the role that ChatGPT can play in supporting and leading the architectural activities of ACSE. Preliminary results of this study demonstrate that ChatGPT is capable of mimicking the role of an architect in ACSE.

Ozkaya et al.~\cite{10109345} present a wide range of potential applications of LLMs in various SE activities including requirements documentation and specification generation, arguing that LLMs can assist in generating more complete specifications significantly quicker. 

Zhang et al.~\cite{zhang2023preliminary} empirically evaluate how ChatGPT performs on requirements analysis tasks to derive insights into how generative large language models, represented by ChatGPT, influence the research and practice of NLP4RE. The evaluation results demonstrate ChatGPT's impressive ability to retrieve requirements information from different types of artifacts involving multiple languages under a zero-shot setting. It is worthwhile for the research and industry communities to study generative large language models in NL RE tasks. 

We can observe the growing interest among the research community to investigate the potential applications of large language models like ChatGPT within various fields of SE including RE, as evidenced by the increasing number of related studies, all within a short period of time. 

\section{Methodology} \label{sec:methods}
\subsection{Research Design}
\begin{figure}[ht!]
  \centering
  \includegraphics[width=1.0\columnwidth]{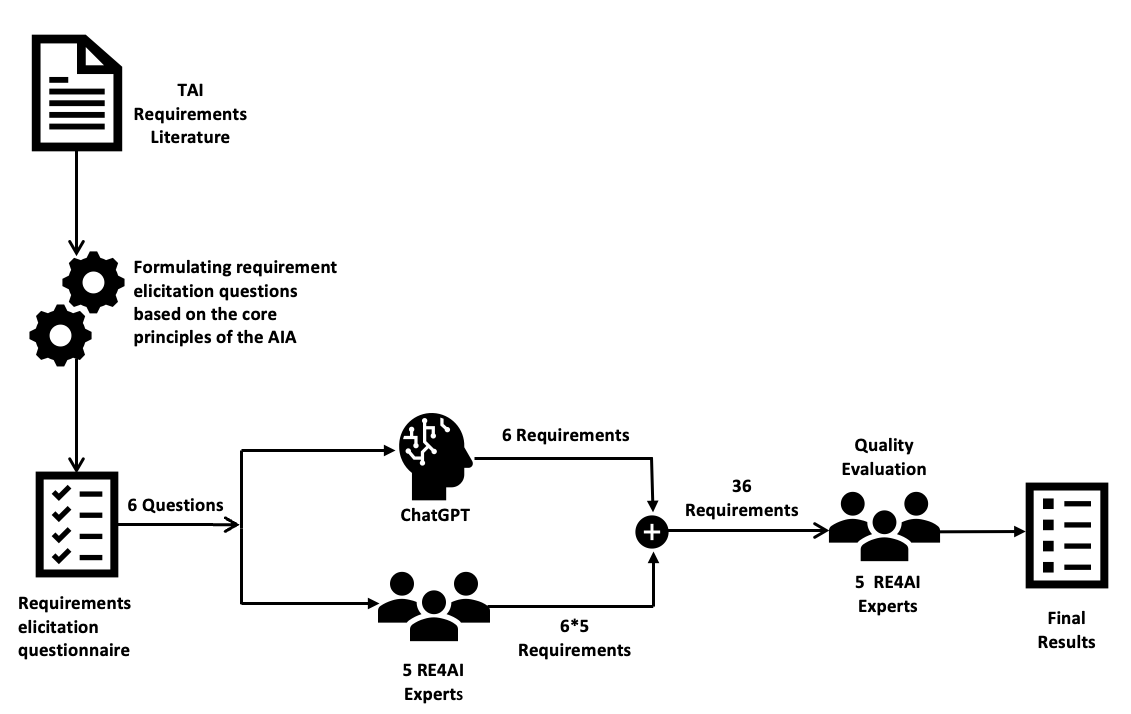}
  \caption{Overview of the methodology used.}
  \label{fig:Methodology}
\end{figure}

We employed a two-stage process to address the research questions as shown in Figure \ref{fig:Methodology}. The first one is using ChatGPT to collect synthetic data, which, in this case, are requirements for developing Trustworthy AI systems. The second method is conducting interview-based surveys. We conducted two rounds of interview-based surveys with different research objectives. The first round was conducted to gather requirements for developing Trustworthy AI systems. These requirements were intended to be very general and not for a particular system. The collected responses included requirements relevant to a wide range of AI and AI-based systems including generative AI models, autonomous vehicles, self-driving cars, AI chatbots, etc. -- all aiming to ensure the system being developed and deployed is Trustworthy. Neither ChatGPT nor the interview participants were provided with any definitions for the Trustworthy AI qualities for the round-1 interview-based survey. The second round was conducted to evaluate the quality of the requirements collected. 

\subsection{Generating Requirements for Trustworthy AI Using ChatGPT} \label{subsec:reqgen}

\begin{figure}[ht!]
  \centering
  \includegraphics[width=1.0\columnwidth]{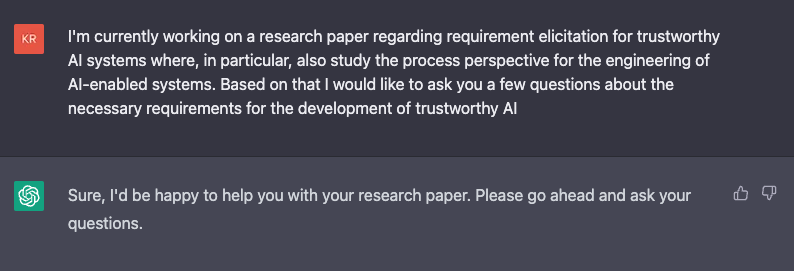}
  \caption{Context Prompt}
  \label{fig:QC}
\end{figure}

This part of the study includes multiple steps. The first step was to formulate 6 questions to create a requirements elicitation questionnaire for developing Trustworthy AI systems. Before asking ChatGPT the 6 questions, it was given a \textit{context prompt} as a conversation kick-off providing some context of what we are trying to achieve as shown in Figure \ref{fig:QC}. The same text in the prompt was mentioned to the participants of the interview-based surveys to ensure all sources from which we were eliciting requirements had the same context.

Then, the 6 questions we crafted were fed to the ChatGPT as inputs. It should be noted that the responses generated by ChatGPT were recorded without any modifications, and screenshots of the responses were taken to ensure the veracity of the data. The requirements elicitation questionnaire is as follows:

\begin{itemize} \label{list:rtaiqs}
   \item \textbf{Q1:} What are the necessary requirements for developing an AI system that ensures its Accuracy and Robustness?
   \item \textbf{Q2:} How to ensure that the data used in training, testing, and validating an AI model is unbiased and fair?
   \item \textbf{Q3:} What requirements are important to ensure the AI model is transparent and explainable?
   \item \textbf{Q4:} What are the most important privacy and security requirements that AI developers need to consider?
   \item \textbf{Q5:} What kind of human oversight requirements need to be in place while developing AI systems to ensure that the AI is Trustworthy?
   \item \textbf{Q6:} Are there any other requirements you can think of that need to be considered while developing AI systems to ensure that the AI is Trustworthy?
\end{itemize}

\textbf{Q1} is crafted to elicit requirements relevant to Accuracy \& Robustness quality property of the Trustworthy AI. Similarly, \textbf{Q2} is for Non-discrimination, \textbf{Q3} is for Transparency \& Explainability, \textbf{Q4} is for Privacy \& Security, \textbf{Q5} is for Human Agency, and \textbf{Q6} is for any other requirements that are relevant to developing a Trustworthy AI system. However, not all of the Trustworthy AI qualities were explicitly included in curating the list of questions presented to the interviewees and ChatGPT. No questions were formulated and asked of either ChatGPT or the interviewees for qualities like Safety, Accountability, Regulations, and Human Agency. This was a deliberate decision to assess the level of awareness of the interviewees and the ChatGPT model regarding these crucial factors related to the requirements of Trustworthy AI systems. By omitting these AI qualities, the study aimed to determine whether the interviewees and ChatGPT would mention them in response to \textbf{Q6} of the requirements elicitation questionnaire, unsolicited. The intention behind this was to gain a deeper understanding of the degree to which both humans and ChatGPT are cognizant of the key AI qualities that are necessary for developing a niche system. The responses generated by ChatGPT for these 6 questions are provided as part of supplementary material in Section \ref{sec:supmat}.

\subsection{Round-1 Interview-based Surveys: Eliciting Requirements for Trustworthy AI} \label{subsec:int1}

To effectively gather the necessary requirements, interviewing individuals with expert knowledge in the field of RE for AI/ML systems (whom we will refer to as expert respondents for the rest of the paper) was believed to be the most suitable approach in our study design. Interview-based surveys with 5 expert respondents were conducted to elicit requirements for developing Trustworthy AI. These expert respondents were selected based on their experience in the field of RE4AI and their familiarity with the concept of Trustworthy AI. The pre-interview formalities included explicitly informing the participants about the purpose of collecting requirements from them, i.e., evaluating the quality of the requirements and publishing the results as a part of a research study. We also informed and took consent from each of the participants to record the interviews, remove any personal identifiers from the interview transcripts to comply with GDPR regulations, and store the recordings and the anonymized transcripts securely along with a master file containing their contact information. It was made clear that their participation is completely voluntary and they can refuse to answer a particular question or end the interview at their discretion. The recording and transcription of the interviews began only after making sure all the participants understood the formalities and gave their consent. The same 6 questions that were given as inputs to the ChatGPT were asked of the expert respondents. Once the planned interview-based surveys were finished and the 30 responses from the expert respondents were obtained, a total of 36 responses, each consisting of a varying number of requirements related to the development of Trustworthy AI, were ready to be evaluated using selected requirement quality attributes. The supplementary material provided in Section \ref{sec:supmat} does not contain round-1 interview participants' responses because the pre-interview informed consent did not involve making their responses publicly accessible.

\subsection{Round-2 Interview-based Surveys: Evaluating the Quality of the Elicited Requirements} \label{subsec:int2}

These 36 responses consisting of requirements for Trustworthy AI were presented to a different set of 5 RE experts (whom we will refer to as expert evaluators for the rest of the paper) for evaluation. These 5 people were chosen based on their experience in working with requirements for AI systems. To avoid any potential bias, the fact that 6 of the responses were generated by ChatGPT was not disclosed to the expert evaluators. The demographic data of the expert respondents was also not revealed. The 36 responses were presented to these expert evaluators in a randomised order. Expert evaluators were also provided with supplementary material with the definitions of Trustworthy AI qualities and the requirements quality attributes before the interview began. This was done in order to ensure a fair and consistent evaluation practice. 

The expert evaluators were then requested to provide a score for each response across seven chosen quality attributes, i.e., Abstraction, Atomicity, Consistency, Correctness, Unambiguity, Understandability, and Feasibility. Each response was rated on a scale of 0-10 for each attribute, with 0 being the lowest score assigned for the poorest quality requirements and 10 being the highest score assigned for the best quality requirements, resulting in quantified measures of the selected requirements quality attributes.


\section{Results \& Analysis} \label{sec:results}

In this section, we present the outcomes of our research study. We report the quality scores for requirements elicited using ChatGPT and compare them with those formulated by expert respondents. The average scores and standard deviations for each requirements quality attribute were computed separately for both the expert respondents' formulated and ChatGPT-generated requirements using the data from round-2 interview-based surveys. We also discuss the shortcomings of using ChatGPT to elicit requirements based on the analysis of our findings in subsection \ref{subsec:int2}. 

\subsection{Quality Scores for ChatGPT-generated Requirements} \label{subsec:rq1}
\begin{figure}[ht!]
   \centering
   \includegraphics[width=0.97\columnwidth]{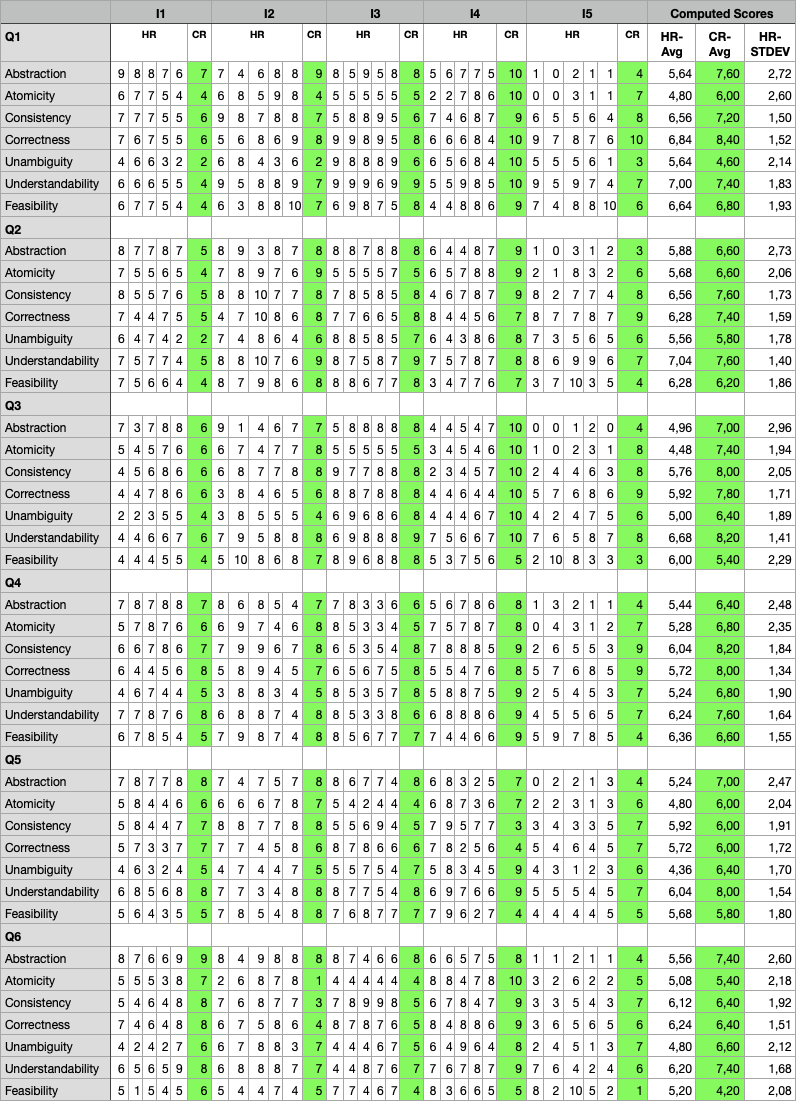}
   \caption{Quality evaluation scores of Trustworthy AI requirements.}
   \label{fig:fig-3}
\end{figure}

\begin{figure}[ht!]
   \centering
   \includegraphics[width=1.0\columnwidth]{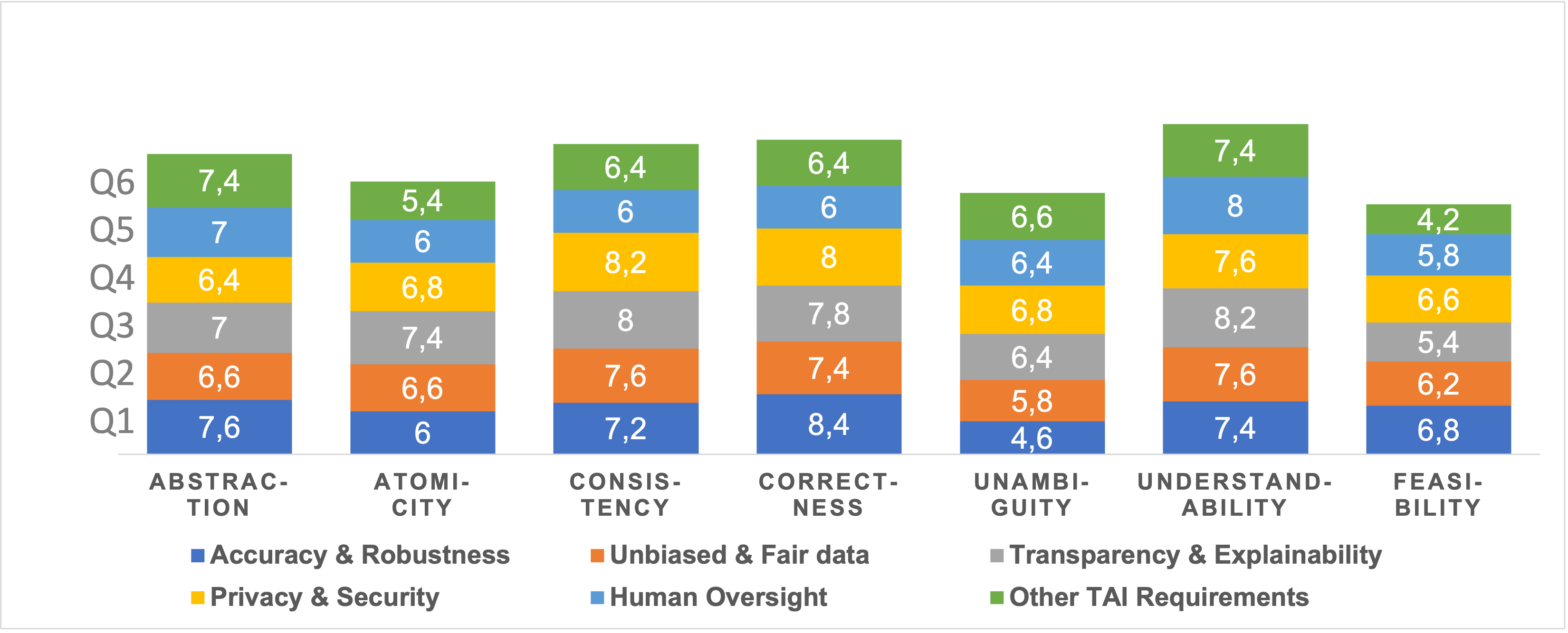}
   \caption{Requirements quality scores of ChatGPT-generated requirements}
   \label{fig:fig-4}
\end{figure}

Figure \ref{fig:fig-3} presents the scores of all 36 responses for each of the 6 questions from \ref{list:rtaiqs} over the 7 requirements quality attributes. I1-I5 refer to the 5 interviews we conducted in round-2 interview-based surveys. Q1-Q6 are the questions given to the expert respondents of round-1 interviews. The columns labelled HR consists of the scores assigned for the requirements provided by the expert evaluators. Columns labelled CR consist of scores assigned for ChatGPT responses (highlighted in green). For example, the first column and row value (9) is the expert evaluator's numerical score of the abstraction quality for the response from the first expert respondent. Highlighted in green, (7) in that same row is the evaluation of the level of Abstraction of ChatGPT's response to the same question. HR-avg is the computed average score for expert evaluators' scores across all 5 round-2 interview-based surveys and across all five responses while CR-avg is the computed average score for ChatGPT-generated responses for the same. HR-STDEV is the computed standard deviation of scores for expert evaluators' scores across all 5 round-2 interview-based surveys. Based on the HR-avg, CR-avg and HR-STDEV presented in Figure \ref{fig:fig-3}, we generated Figure \ref{fig:fig-4}, which is used as a basis to answer \textbf{RQ1} and \textbf{RQ2}.

Figure \ref{fig:fig-4} provides a comparison of the quality of the requirements generated by ChatGPT for different Trustworthy AI qualities over the 7 selected requirements quality attributes. The first column represents Abstraction quality scores for all 6 questions from \ref{list:rtaiqs} with blue layer answering \textbf{Q1}, orange layer answering \textbf{Q2}, grey layer answering \textbf{Q3}, yellow layer answering \textbf{Q4}, light blue layer answering \textbf{Q5}, and green layer answering \textbf{Q6}. Similarly, the second column represents Atomicity quality scores, the third column represents Consistency quality scores, the fourth column represents Correctness quality scores, the fifth column represents Unambiguity quality scores, the sixth column represents Understandability quality scores and the seventh column represents Feasibility quality scores. 

The overall height of the stacked bar represents the total score achieved by ChatGPT-generated requirements for each requirement quality parameter across all Trustworthy AI qualities. As observed in Figure \ref{fig:fig-4}, the tallest stacked bar is the Understandability column, i.e., the most attested attribute of ChatGPT-generated requirements is Understandability. This is closely followed by Correctness, Consistency and Abstraction. Unabmiguity and Feasibility are the second shortest and the shortest stacked bars respectively, i.e., they are the least attested quality attributes of ChatGPT-generated requirements.

The results indicated that Correctness with an average score of 8.4 out of 10, and Abstraction with an average score of 7.6, were the most prominent qualities of the requirements for ensuring Accuracy \& Robustness, while Unambiguity was the least impressive quality, with an average score of only 4.6.

Coming to requirements or ensuring the usage of Unbiased \& Fair data in training, testing \& validating an AI model, Consistency and Understandability were jointly ranked the highest, with an average score of 7.6, followed closely by Correctness with an average score of 7.4. However, Unambiguity scored the least in this aspect with an average score of 5.8.

For Transparency \& Explainability requirements, Understandability was the highest-ranked quality with an average score of 8.2, followed by Consistency and Correctness, with scores of 8 and 7.8 respectively. Feasibility scored the lowest, with an average score of 5.4. 

Regarding Privacy \& Security requirements, Consistency was the most highly rated quality, with an average score of 8.2 while Abstraction scored the lowest, with an average score of 6.4.

For Human Oversight requirements, Understandability scored the highest with an average score of 8, while Unambiguity was the least impressive, with an average score of 5.4. Finally, the requirements generated for \textbf{Q6} of the interview questionnaire had Abstraction and Understandability as their most attested quality, with an average score of 7.4, while Feasibility scored the least, an unimpressive 4.2. 

\subsection{Shortcomings of ChatGPT in Generating Requirements} \label{subsec:rq2}

From Figure \ref{fig:fig-4} and the results presented in Section \ref{subsec:rq1}, it is observed that the requirements generated by ChatGPT exhibited high levels of Understandability, Consistency, and Correctness, which are essential qualities for good requirements. ChatGPT-generated requirements achieved good scores for Abstraction for most Trustworthy AI principles except for Privacy \& Security. On the other hand, Unambiguity and Feasibility achieved the lowest scores, making them the least prominent quality attributes in ChatGPT's requirements.

Low Unambiguity of requirements could mean that presentation of the requirements is unclear and imprecise and could also lead to multiple interpretations. Low Feasibility indicates that the requirements cannot be realistically implemented. This could be due to technological constraints (unavailability of the recommended infrastructure or access to the recommended tech stack), resource limitations (time, budget, manpower) or dependency on external factors beyond the project stakeholders' control. Although no effort was made to investigate the precise cause of the low Unambiguity and Feasibility scores of ChatGPT-generated requirements, it can be concluded that the requirements are not clear and feasible enough to implement in real-world projects yet. 

Additionally, ChatGPT's response to \textbf{Q6} included requirements related to Human Rights, Ethical Considerations, Interoperability, Sustainability, Responsible Usage and Stakeholder Diversity during the development of the AI systems. But we did not find any requirements related to Accountability, Regulations, or Safety. In comparison, the requirements provided by human RE4AI experts  for \textbf{Q6} were centred around Sustainability, Sovereignty, User Experience, Safety, Human Factors and System Predictability.

\subsection{Comparing ChatGPT-generated Requirements with Human Expert Requirements} \label{subsec:rq3}

\begin{figure}[ht!]
   \centering
   \includegraphics[width=1.0\columnwidth]{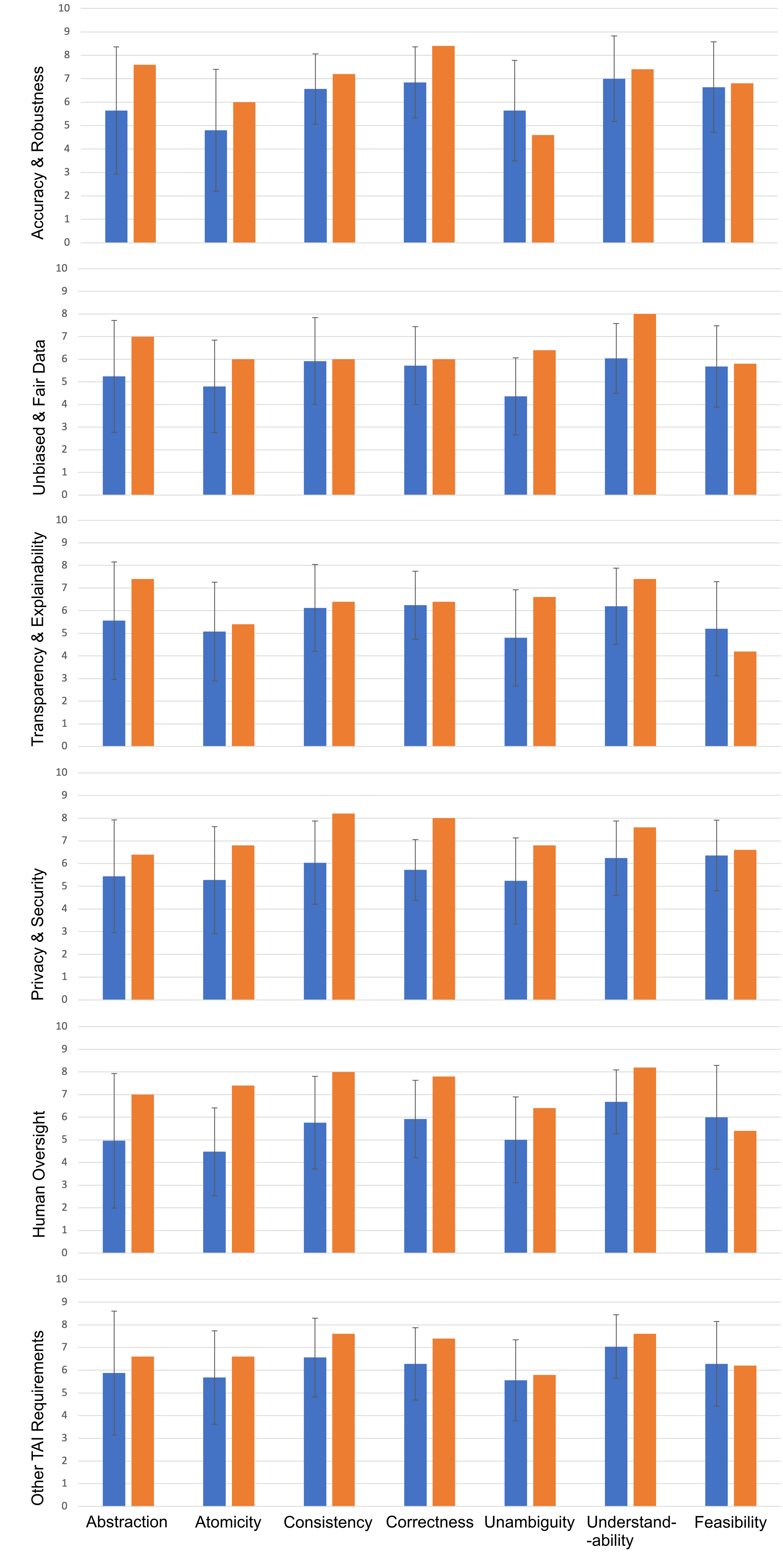}
   \caption{Comparing ChatGPT generated requirements with Human formulated requirements. The blue bar represents the average scores of the requirements formulated by Human RE4AI experts while the orange bar represents the average scores of the requirements generated by ChatGPT. The whisker within the blue bar represents the computed standard deviation.}
   \label{fig:fig-5}
\end{figure}

From Figure \ref{fig:fig-5}, we can see that the orange bar representing the average scores of the ChatGPT-generated requirements is taller than the blue bar, which represents the average score of human-formulated requirements in most instances. It means quality evaluation scores of ChatGPT-generated requirements outperform the scores achieved by expert respondents' formulated requirements in most cases. The orange bar is shorter than the blue bar in only 4 instances; 1) Unambiguity of Accuracy \& Robustness requirements, 2) Feasibility of Transparency \& Explainability requirements, 3) Feasibility of Human Oversight requirements, and 4) Feasibility of other Trustworthy AI requirements. Once again, Feasibility and Unambiguity are the only attributes where the average scores of expert respondents' formulated requirements outperformed the ChatGPT-generated requirements, in 4 out of 42 instances.


\section{Discussions} \label{sec:discussion}

Based on our findings presented in Section \ref{sec:results}, we observe that ChatGPT-generated requirements are considered to be acceptable by RE experts (expert evaluators) in direct competition with expert respondents' formulated requirements on multiple quality attributes. But this does not mean they are flawless. Generating ambiguous and unfeasible requirements is a clear shortcoming identified in using ChatGPT to elicit requirements, even if the average scores of ChatGPT-generated requirements for Unambiguity and Feasibility are still higher than human-formulated requirements in the majority of instances. Feasibility is especially important to consider as a requirement is only of value if it can be transformed into a design and an implementation with reasonable effort and cost~\cite{denger2005quality}. Ambiguous presentation of requirements may lead to uncertainty in the decision-making process during the design stage, which is considered highly undesirable by developers~\cite{mu2005inconsistency}.

Requirements for a particular system to be developed should come from or be approved by the customer. This is a key element of RE that cannot (and perhaps should not) be replaced by AI. However, AI can assist to some extent in various RE tasks that are tedious yet require complex reasoning abilities. Based on our results, we find that LLMs can assist requirements analysts in at least making requirements more Abstract, Atomic, Complete, and Understandable.

An example use case may look as follows: A product owner receives the requirements from the customer(s) through interviews or focus groups while ensuring there is a textual transcript of the entire conversation. This transcript can be refined using a LLM to convert it into a system requirements specification (SRS) which has highly Abstract (if required), Atomic, Consistent, Correct and easily Understandable set of requirements. This process can reduce time and effort compared to preparing requirements manually. In this way, the adoption of LLMs in the RE process can lead to increased process efficiency since analysts can redirect their efforts to activities that require advanced critical thinking. In addition, outputs from LLMs may have improved quality compared to manually crafted outputs, as observed here. 

The level of awareness regarding high-level requirements that an AI system must adhere to be deemed Trustworthy is critical for the successful development of Trustworthy AI systems. However, of the five interviewees, only one identified the significance of incorporating Safety and Human Agency requirements into Trustworthy AI development as a response to \textbf{Q6} from the requirements elicitation questionnaire. None of the other interviewees provided any requirements related to Accountability, Regulations, or Safety and neither did ChatGPT. This highlights the significance of possessing domain knowledge when formulating Trustworthy AI requirements, irrespective of the method that is being employed to elicit and formulate requirements. It holds true even while critically reflecting on ChatGPT-supplied content for the RE process to ensure the Accuracy of the output. But this should not necessarily overshadow the promise ChatGPT showed in providing content that is considered to be Abstract, Consistent, Correct and Understandable requirements by RE4AI experts. Instead, we believe this promise should be fostered by trying to address the shortcomings.

But there is a bigger challenge to overcome for the usage of LLMs like ChatGPT to support the requirements elicitation process, namely, hallucinations. The model is prone to generating factual-sounding statements that cannot be validated from the source, a phenomenon referred to as extrinsic hallucination~\cite{bang2023multitask}. Despite scoring higher on the Correctness attribute, and having no evidence of the ChatGPT's output in our study being affected by the hallucination effect, implementing ChatGPT-generated requirements into the AI systems development process without any Human Oversight mechanisms in place might lead to an edge-case scenario where a factually incorrect requirement might be perceived as a correct one. 

Recent advancements in LLMs have shown indications of efforts to address these limitations. The development of GPT-4  has shown significant improvement over existing models in various NLP tasks. In particular, GPT-4 improves over the latest GPT-3.5 model, on which ChatGPT is based, by 19 percentage points, with significant gains across all topics, and has surpassed the majority of state-of-the-art systems, which typically require task-specific fine-tuning. This advancement is attributable to predictable scaling, which has enhanced measures of factuality and adherence to desired behaviour, demonstrating the potential of GPT-4 to offer improved language understanding capabilities and to facilitate the development of more advanced NLP systems~\cite{openai2023gpt4}.

Apart from that, recent studies have also demonstrated that LLMs exhibit remarkable performance gains when trained using a small amount of in-context data using one-shot and few-shot prompting techniques~\cite{brown2020language}, with the performance being heavily influenced by the domain of the corpus source~\cite{shin2022effect}. It has also been observed that transformer models, which have been pre-trained with software engineering (SE) domain data, exhibit superior performance on SE-related applications as compared to general domain models and can be regarded as the current state-of-the-art for SE use cases~\cite{von2022validity}. Further research has shown that fine-tuning such models on smaller datasets in combination with transfer learning can significantly enhance their performance on specific tasks with limited data~\cite{kasneci2023chatgpt}. 

This multitude of available approaches, if leveraged, could help the research community to address the identified challenges and come up with robust mitigation strategies, fostering further research into investigating ways to utilize LLMs in RE activities. The findings presented in this study can be viewed as preliminary proof of concept, which could provide motivation for further research to explore and evaluate the boundary of robust state-of-the-art LLMs application in RE activities with appropriate Human Oversight mechanisms in place to ensure the ethical and responsible application of these technologies.

\subsection{Future Work}
One potential avenue for future research is to conduct more rigorous and comprehensive evaluation studies that involve a wider and more diverse range of factors like utilizing the latest version of the GPT architecture, such as GPT-4, over multiple system domains along with a variety of prompt engineering techniques to overcome the identified limitations. 

To enhance the LLM's interpretation of the question and generate more accurate responses, users can improve the situation by managing the context of the dialogue \cite{white2023prompt}. One potential direction is to observe the effect of providing varying degrees of context on the output while interacting with the LLM. For example, giving information about the system's goal, the intended users and the development environment might result in more concrete/unambiguous and feasible requirements. 

Another promising approach could be to use Knowledge Graphs (KG) to enhance the factuality of LLM's output. KG-enhanced LLM inference utilizes KGs during the inference stage of LLMs, which enables LLMs to access the latest knowledge without retraining~\cite{pan2023unifying}.

\subsection{Threats to Validity} \label{sec:threats}
We followed Runeson \& H\"{o}st's guidelines for conducting qualitative research analysis~\cite{runeson2009guidelines} in software engineering to discuss any possible threats to the validity of this research study. One of the possible threats to the validity of this study is the generalizability of the results. Since the requirements were gathered for Trustworthy AI systems development, which in itself is a rapidly evolving domain, these results might or might not hold for systems from other domains as well. 

Another threat to validity would be the construct type. Neither ChatGPT nor the requirements elicitation interview participants were provided definitions for Accuracy \& Robustness, Safety, Non-discrimination, Transparency \& Explainability, Accountability, Privacy \& Security, Regulations, and Human Agency \& Oversight. This was intentionally done to encourage the responses and requirements to be influenced by the domain knowledge of the participants and ChatGPT. 

For computational simplicity, the scores provided by expert evaluators were averaged out. This could potentially be a threat to validity as the variation in the human expert responses is minimized. Nevertheless, the standard deviation of human expert responses (HR-STDEV) was provided in Figure \ref{fig:fig-3} to represent the variety in the experts' opinions.

Since the expectation was to receive one-shot answers to the questions, the limitations observed may be a result of the constrained interaction employed for querying ChatGPT, as the chat is designed to facilitate continued interaction. 


\section{Conclusion} \label{sec:conclusion}

In conclusion, this study aimed to explore the potential of ChatGPT in eliciting requirements and comparing its output with requirements formulated by five RE4AI experts from academia and industry. The quality of requirements was evaluated by interviewing an additional five RE4AI experts in our study. Results from our experiment show that ChatGPT-generated requirements are considered highly Abstract, Atomic, Consistent, Correct, and Understandable in comparison to human RE experts' formulated requirements. Unambiguity and Feasibility of the requirements received lower scores in comparison to scores of other requirements' quality attributes. Our findings suggest that ChatGPT has promising potential to support requirements elicitation processes, like converting raw requirements documents into high-quality specification documents, ensuring consistency, and improving Understandability among other things. ChatGPT's use cases should be further investigated in various RE activities to leverage the emergent behaviour of LLMs more effectively and foster wider adoption of LLMs in natural language-based RE activities.

\section{Acknowledgement}
We want to thank Dr.~Beatriz Cabrero-Daniel for her encouragement \& valuable support. This work was supported by the Vinnova project ASPECT [2021-04347].

\section{Supplementary Material} \label{sec:supmat}
\url{https://doi.org/10.5281/zenodo.8124936}. 

\bibliographystyle{plain}
\bibliography{main}

\begin{thebibliography}{10}

\bibitem{AIact}
{\em EUR-Lex - 52021PC0206 - EN - EUR-Lex}.

\bibitem{10062688}
{Abdullah, Malak and Madain, Alia and Jararweh, Yaser}.
\newblock {ChatGPT: Fundamentals, Applications and Social Impacts}.
\newblock In {\em {2022 Ninth International Conference on Social Networks
  Analysis, Management and Security (SNAMS)}}, pages 1--8, 2022.

\bibitem{ahmad2023towards}
{Ahmad, Aakash and Waseem, Muhammad and Liang, Peng and Fehmideh, Mahdi and
  Aktar, Mst Shamima and Mikkonen, Tommi}.
\newblock {Towards Human-Bot Collaborative Software Architecting with ChatGPT}.
\newblock {\em arXiv preprint arXiv:2302.14600}, 2023.

\bibitem{alhoshan2023zero}
{Alhoshan, Waad and Ferrari, Alessio and Zhao, Liping}.
\newblock {Zero-shot Learning for Requirements Classification: An Exploratory
  Study}.
\newblock {\em {Information and Software Technology}}, page 107202, 2023.

\bibitem{bang2023multitask}
{Bang, Yejin and Cahyawijaya, Samuel and Lee, Nayeon and Dai, Wenliang and Su,
  Dan and Wilie, Bryan and Lovenia, Holy and Ji, Ziwei and Yu, Tiezheng and
  Chung, Willy}.
\newblock {A Multitask, Multilingual, Multimodal Evaluation of ChatGPT on
  Reasoning, Hallucination, and Interactivity}.
\newblock {\em arXiv preprint arXiv:2302.04023}, 2023.

\bibitem{brown2020language}
{Brown, Tom and Mann, Benjamin and Ryder, Nick and Subbiah, Melanie and Kaplan,
  Jared D and Dhariwal, Prafulla and Neelakantan, Arvind and Shyam, Pranav and
  Sastry, Girish and Askell, Amanda and others}.
\newblock {Language Models are Few-shot Learners}.
\newblock {\em {Advances in Neural Information Processing Systems}},
  33:1877--1901, 2020.

\bibitem{choi2023chatgpt}
{Choi, Jonathan H and Hickman, Kristin E and Monahan, Amy and Schwarcz,
  Daniel}.
\newblock {ChatGPT Goes to Law School}.
\newblock {\em {Available at SSRN}}, 2023.

\bibitem{9218169}
{Dechesne, Francien}.
\newblock {Requirements Engineering for Moral Considerations in Algorithmic
  Systems : RE’20 Conference Keynote}.
\newblock In {\em {2020 IEEE 28th International Requirements Engineering
  Conference (RE)}}, pages 1--2, 2020.

\bibitem{denger2005quality}
{Denger, Christian and Olsson, Thomas}.
\newblock {Quality Assurance in Requirements Engineering}.
\newblock In {\em {Engineering and Managing Software Requirements}}, pages
  163--185. Springer, 2005.

\bibitem{genova2013framework}
{G{\'e}nova, Gonzalo and Fuentes, Jos{\'e} M and Llorens, Juan and Hurtado,
  Omar and Moreno, Valent{\'\i}n}.
\newblock {A framework to Measure and Improve the Quality of Textual
  Requirements}.
\newblock {\em {Requirements engineering}}, 18(1):25--41, 2013.

\bibitem{haque2022think}
{Haque, Mubin Ul and Dharmadasa, Isuru and Sworna, Zarrin Tasnim and Rajapakse,
  Roshan Namal and Ahmad, Hussain}.
\newblock {" I think this is the most disruptive technology": Exploring
  Sentiments of ChatGPT Early Adopters using Twitter Data}.
\newblock {\em arXiv preprint arXiv:2212.05856}, 2022.

\bibitem{aihleg}
{High-Level Expert Group on Artificial Intelligence}.
\newblock {\em {Ethics Guidelines for Trustworthy Artificial Intelligence
  (AI)}}.
\newblock 2019.

\bibitem{iyer2021ai}
{Iyer, Lakshmi Shankar}.
\newblock {AI-Enabled Applications Towards Intelligent Transportation}.
\newblock {\em {Transportation Engineering}}, 5:100083, 2021.

\bibitem{ji2023survey}
{Ji, Ziwei and Lee, Nayeon and Frieske, Rita and Yu, Tiezheng and Su, Dan and
  Xu, Yan and Ishii, Etsuko and Bang, Ye Jin and Madotto, Andrea and Fung,
  Pascale}.
\newblock {Survey of Hallucination in Natural Language Generation}.
\newblock {\em {ACM Computing Surveys}}, 55(12):1--38, 2023.

\bibitem{white2023prompt}
{Jules White and Quchen Fu and Sam Hays and Michael Sandborn and Carlos Olea
  and Henry Gilbert and Ashraf Elnashar and Jesse Spencer-Smith and Douglas C.
  Schmidt}.
\newblock {A Prompt Pattern Catalog to Enhance Prompt Engineering with
  ChatGPT}, 2023.

\bibitem{kasneci2023chatgpt}
{Kasneci, Enkelejda and Se{\ss}ler, Kathrin and K{\"u}chemann, Stefan and
  Bannert, Maria and Dementieva, Daryna and Fischer, Frank and Gasser, Urs and
  Groh, Georg and G{\"u}nnemann, Stephan and H{\"u}llermeier, Eyke}.
\newblock {ChatGPT for Good? On Opportunities and Challenges of Large Language
  Models for Education}.
\newblock {\em {Learning and Individual Differences}}, 103:102274, 2023.

\bibitem{kaur2021requirements}
{Kaur, Davinder and Uslu, Suleyman and Durresi, Arjan}.
\newblock {Requirements for Trustworthy Artificial Intelligence - A Review}.
\newblock In {\em {Advances in Networked-Based Information Systems: The 23rd
  International Conference on Network-Based Information Systems (NBiS-2020)
  23}}, pages 105--115. Springer, 2021.

\bibitem{kaur2022Trustworthy}
{Kaur, Davinder and Uslu, Suleyman and Rittichier, Kaley J and Durresi, Arjan}.
\newblock {Trustworthy Artificial Intelligence: A Review}.
\newblock {\em {ACM Computing Surveys (CSUR)}}, 55(2):1--38, 2022.

\bibitem{khurana2022natural}
{Khurana, Diksha and Koli, Aditya and Khatter, Kiran and Singh, Sukhdev}.
\newblock {Natural Language Processing: State-of-the-Art, Current Trends and
  Challenges}.
\newblock {\em {Multimedia Tools and Applications}}, pages 1--32, 2022.

\bibitem{kostova2020interplay}
{Kostova, Blagovesta and Gurses, Seda and Wegmann, Alain}.
\newblock {On the Interplay between Requirements, Engineering, and Artificial
  Intelligence.}
\newblock In {\em REFSQ Workshops}, 2020.

\bibitem{lee2019confident}
{Lee, Hosub and Kobsa, Alfred}.
\newblock {Confident Privacy Decision-making in IoT Environments}.
\newblock {\em {ACM Transactions on Computer-Human Interaction (TOCHI)}},
  27(1):1--39, 2019.

\bibitem{mu2005inconsistency}
{Mu, Kedian and Jin, Zhi and Lu, Ruqian}.
\newblock {Inconsistency-based Strategy for Clarifying Vague Software
  Requirements}.
\newblock In {\em {Australian Conference on Artificial Intelligence}}, pages
  39--48. Springer, 2005.

\bibitem{openai2023gpt4}
{OpenAI}.
\newblock {GPT-4 Technical Report}, 2023.

\bibitem{10109345}
Ipek Ozkaya.
\newblock {Application of Large Language Models to Software Engineering Tasks:
  Opportunities, Risks, and Implications}.
\newblock {\em {IEEE Software}}, 40(3):4--8, 2023.

\bibitem{runeson2009guidelines}
{Runeson, Per and H{\"o}st, Martin}.
\newblock {Guidelines for Conducting and Reporting Case Study Research in
  Software Engineering}.
\newblock {\em {Empirical Software Engineering}}, 14:131--164, 2009.

\bibitem{shen2023chatgpt}
{Shen, Yiqiu and Heacock, Laura and Elias, Jonathan and Hentel, Keith D and
  Reig, Beatriu and Shih, George and Moy, Linda}.
\newblock {ChatGPT and Other Large Language Models Are Double-edged Swords},
  2023.

\bibitem{shin2022effect}
{Shin, Seongjin and Lee, Sang-Woo and Ahn, Hwijeen and Kim, Sungdong and Kim,
  HyoungSeok and Kim, Boseop and Cho, Kyunghyun and Lee, Gichang and Park,
  Woomyoung and Ha, Jung-Woo}.
\newblock {On the Effect of Pre-training Corpora on In-context Learning by A
  Large-scale Language Model}.
\newblock {\em arXiv preprint arXiv:2204.13509}, 2022.

\bibitem{pan2023unifying}
year={2023} {Shirui Pan and Linhao Luo and Yufei Wang and Chen Chen and Jiapu
  Wang and Xindong Wu}.
\newblock {Unifying Large Language Models and Knowledge Graphs: A Roadmap}.

\bibitem{standards2016ieee}
{Standards Association and others}.
\newblock {IEEE Recommended Practice for Software Requirements Specifications.
  IEEE Std 830 1998}, 2016.

\bibitem{von2022validity}
{Von der Mosel, Julian and Trautsch, Alexander and Herbold, Steffen}.
\newblock {On the Validity of Pre-trained Transformers for Natural Language
  Processing in the Software Engineering Domain}.
\newblock {\em {IEEE Transactions on Software Engineering}}, 2022.

\bibitem{zhang2022would}
{Zhang, Bowen and Ding, Daijun and Jing, Liwen}.
\newblock {How would Stance Detection Techniques Evolve after the Launch of
  ChatGPT?}
\newblock {\em arXiv preprint arXiv:2212.14548}, 2022.

\bibitem{zhang2023preliminary}
{Zhang, Jianzhang and Chen, Yiyang and Niu, Nan and Liu, Chuang}.
\newblock {A Preliminary Evaluation of ChatGPT in Requirements Information
  Retrieval}.
\newblock {\em arXiv preprint arXiv:2304.12562}, 2023.

\bibitem{zhao2021natural}
{Zhao, Liping and Alhoshan, Waad and Ferrari, Alessio and Letsholo, Keletso J
  and Ajagbe, Muideen A and Chioasca, Erol-Valeriu and Batista-Navarro, Riza
  T}.
\newblock {Natural Language Processing for Requirements Engineering: A
  Systematic Mapping Study}.
\newblock {\em {ACM Computing Surveys (CSUR)}}, 54(3):1--41, 2021.

\end{thebibliography}

\end{document}